\documentclass[aps,pre,amsmath,amssymb,twocolumn,floatfix]{revtex4}
\usepackage{graphicx}
\usepackage{epsfig}
\newcounter{abc}

\input epsf

\newcommand{\be}{\begin{equation}}
\newcommand{\ee}{\end{equation}}
\newcommand{\bea}{\begin{eqnarray}}
\newcommand{\eea}{\end{eqnarray}}

\newcommand{\f}{f}

\newcommand{\p}{^{(0)}}

\newcommand{\one}{^{(1)}}
\newcommand{\two}{^{(2)}}

\newcommand{\ronesig}{\rho\one(\sigma)}
\newcommand{\rtwosig}{\rho\two(\sigma)}

\begin{document}

\title{A polydisperse lattice-gas model}
\author{Nigel B. Wilding$^1$}
\author{Peter Sollich$^2$}
\author{Matteo Buzzacchi$^1$}
\affiliation{$1$.~Department of Physics, University of Bath, Bath BA2 7AY, U.K.\\
$2.$~Department of Mathematics, King's College London, Strand, London WC2R 2LS, U.K.}

\begin{abstract} 

We describe a lattice-gas model suitable for studying the generic
effects of polydispersity on liquid-vapor phase equilibria. Using Monte
Carlo simulation methods tailored for the accurate determination of
phase behaviour under conditions of fixed polydispersity, we trace the
cloud and shadow curves for a particular Schulz distribution of the
polydisperse attribute. Although polydispersity enters the model solely
in terms of the strengths of the interparticle interactions, this is
sufficient to induce the broad separation of cloud and shadow curves
seen both in more realistic models and experiments. 

\end{abstract} 
%\pacs{64.70.Jx, 64.60.Fr} 
\maketitle 
\epsfclipon  

\section{Introduction}

Lattice-based models of many-body systems have been in the vanguard of
the drive for an improved understanding of equilibrium phase behaviour,
almost since the inception of the field \cite{CL}. The utility of such
models derives both from the simplifications to approximate analytical
treatments that result from the spatial discretisation, and from their
amenability to efficient numerical simulation. Indeed, even in the
current era of teraflop computers, simulations of lattice models are
still deployed routinely in fields as diverse as polymer physics and
quantum chromodynamics.

The purpose of this paper is to introduce a new lattice-gas model
suitable for simulation investigations of the phase behaviour of {\em
polydisperse fluids}. These are complex fluids whose constituent
particles exhibit continuous variation in terms of some physical
attribute such as their size, shape or charge. Polydispersity is
inherent to a host of natural and synthetic soft matter fluids
including ({\em inter alia}) colloidal dispersions, polymer solutions,
liquid crystals, as well as some biological fluids such as blood
\cite{larson1999}. Understanding its generic effects on the phase
behaviour of model systems is, therefore, not only a matter of
fundamental interest, but also one of considerable practical and
commercial importance. 

The complexity bestowed on a fluid system by polydispersity complicates
simulation studies of phase behaviour. Although newly-developed
methodologies \cite{wilding2002d,wilding2003a,
buzzacchi2006,Wilding2006} allow the accurate study of phase
coexistence in off-lattice models of polydisperse fluids, progress has
nevertheless been hindered (compared to studies of monodisperse
systems) by the fact that coexistence occurs over a {\em region} of the
pressure-temperature phase diagram, rather than merely along a line
\cite{Bellier-castella2002,Rascon2003}. The necessity of exploring a
higher dimensional space of model parameters in order to fully
characterize a given phase transition renders such investigations
laborious. Hence progress in elucidating the {\em generic} effects of
polydispersity on phase behaviour has been slower than be might be
desired.

Accordingly, there is a pressing need for a computationally tractable
model system, which, whilst not necessarily portraying realistically
the microscopic features of any particular polydisperse fluid, does
nevertheless capture principal features of the macroscopic phase
behaviour.  The lattice-gas model presented below should go some way to
meeting this need. Specifically, it should help answer unsolved
questions such as how the coexistence and critical point properties
depend on the nature of the particle interactions and the form and
degree of the polydispersity. Additionally it should provide an
efficient test bed for the development of fresh computational and
analytical methodologies for studying polydisperse phase behaviour.

The structure of our paper is as follows. In Sec.~\ref{sec:bg} we
outline salient features of the phase behaviour of polydisperse
systems, before moving on to describe the polydisperse lattice-gas
(PLG) model in Sec.~\ref{sec:model}. We then report, in
Sec.~\ref{sec:sims}, a simulation study of key features of the
liquid-vapor phase behaviour for one particular choice of the form of
the polydispersity. Finally, Sec.~\ref{sec:concs} summarizes our
findings and conclusions.

\section{Background: polydisperse phase equilibria}
\label{sec:bg}

In this section we present a brief overview of the principal differences
between phase coexistence in polydisperse and monodisperse systems, For
a more detailed discussion, the interested reader is referred to a
recent review \cite{sollich2002}.

The state of a polydisperse system (or any of its phases) is described
by a density distribution $\rho(\sigma)$, with
$\rho(\sigma)d\sigma$ the number density of particles whose
polydisperse attribute $\sigma$ lies in the range $\sigma\ldots
\sigma+d\sigma$. In the most commonly encountered experimental
situation, the form of the overall or ``parent'' distribution,
$\rho\p(\sigma)$, is fixed by the synthesis of the fluid, and only its
scale can vary depending on the proportion of the sample volume
occupied by solvent. One can then write $\rho\p(\sigma)=n\p
f\p(\sigma)$ where $f\p(\sigma)$ is the normalized parent shape
function and $n\p=N/V$ the overall particle number density; as the
latter is varied, $\rho\p(\sigma)$ traces out a ``dilution line'' in
density distribution space. The phase diagram is then spanned by $n\p$
and the temperature $T$.

The richness of the phase behaviour of polydisperse fluids stems from
the occurrence of {\em fractionation}
\cite{evans2001,fairhurst2004,erne2005}: at coexistence, particles of
each $\sigma$ may partition themselves unevenly between two or more
``daughter'' phases as long as--due to particle conservation--the
overall density distribution $\rho\p(\sigma)$ of the parent phase is
maintained. As a consequence, the conventional vapor-liquid binodal of a
monodisperse system splits into a {\em cloud curve} marking the onset
of coexistence, and a {\em shadow curve} giving the density of the
incipient phase; the critical point appears at the intersection of
these curves rather than at the maximum of either~\cite{sollich2002}.
This splitting is seen in experiments on polydisperse fluids (see
e.g.~ref.~\cite{Borchard1994}).

The daughter distributions are related to the parent via a simple
volumetric average: $(1-\xi)\ronesig+\xi\rtwosig=\rho\p(\sigma)\:,$
where $1-\xi$ and $\xi$ are the fractional volumes of the two phases.
In contrast to a monodisperse system where the densities of the
coexisting phases are specified by the binodal and depend solely on
temperature, full specification of the coexistence properties of a
polydisperse system requires not only knowledge of the cloud curve, but
also the dependence of $\xi$, $\ronesig$ and $\rtwosig$ on $n\p$ and
$T$ \cite{buzzacchi2006}.

\section{Polydisperse lattice-gas model}
\label{sec:model}

Our polydisperse lattice-gas (PLG) model is defined within the grand
canonical ensemble by the hamiltonian:

\begin{equation}
H=-\sum_{(ij),\sigma,\sigma^\prime} (\sigma \sigma^\prime)^\alpha c_i(\sigma) c_j(\sigma^\prime)-\sum_{i,\sigma}\mu(\sigma)c_i(\sigma).
\label{eq:hamiltonian}
\end{equation}
Here $\sigma$ is the value (assumed scalar) of the polydisperse
attribute, which controls the strength of interparticle interactions.
We shall regard $\sigma$ as the label for a notional particle
``species'', whose corresponding chemical potential is $\mu(\sigma)$.
The exponent $\alpha$  can, in principle, take an arbitrary value but we
shall restrict ourselves to the case $\alpha=1$ in the present study. $c_i(\sigma)$ simply
counts the number of particles of species $\sigma$ at site $i$, for
which we impose a hard-core constraint such that $\sum_\sigma
c_i(\sigma) = 0$ or $1$. The instantaneous density distribution follows
as $\rho(\sigma)=L^{-d}\sum_ic_i(\sigma)$, with $d=3$ in the
simulations reported below; $i$ runs over the sites of a periodic
lattice $i=1,...,L^{d}$, assumed simple cubic in this work. The sum in
the first term on the right hand side of (\ref{eq:hamiltonian})
similarly runs over all pairs $i$,$j$ of nearest neighbor sites, as
well as over all combinations of $\sigma$ and $\sigma^\prime$. 

Put simply, each lattice site may be either vacant, or occupied by
a single particle which carries a continuous species label $\sigma$.
Particles on nearest neighbor sites interact with a potential energy
which is the negative of the product of their species labels.  Thus the
role of polydispersity in this model is to engender a spread of
possible interaction strengths between particles --a situation which
contrasts with the single interaction strength characterizing the
simple Ising lattice-gas model \cite{BINNEY}. 

The model of Eq.~\ref{eq:hamiltonian} (which we have also briefly
described in a different context elsewhere \cite{buzzacchi2006}) has,
to our knowledge, not been considered previously by other authors. We
note that it is distinct from well-known lattice spin models such as
the $q$-state Potts model in the limit of large $q$, and the $XY$
model. The distinction arises both in terms of the form of the
interactions, and the absence of an isomorphic magnetic description, in
contrast to the case of the simple (monodisperse) lattice-gas \cite{BINNEY}.

\section{Simulation studies}
\label{sec:sims}

\subsection{Parent distribution}
\label{sec:parent}

\begin{figure}
\includegraphics[type=eps,ext=.eps,read=.eps,width=0.9\columnwidth,clip=true]{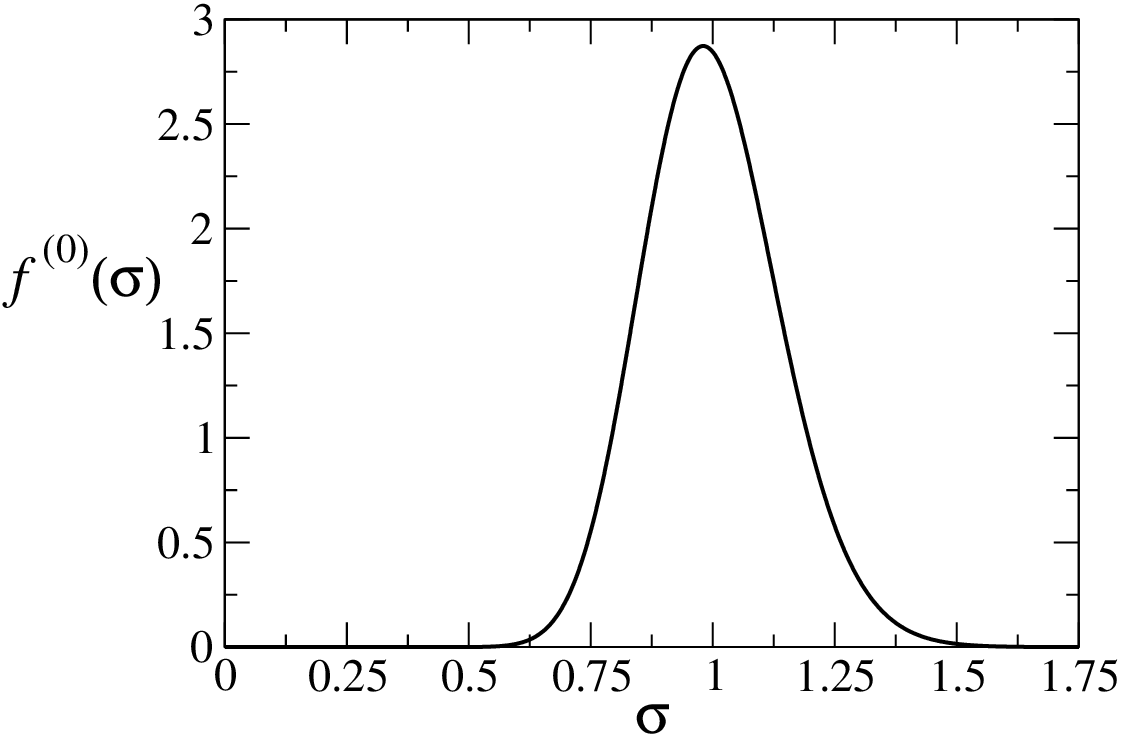}
\caption{The Schulz parent distribution for the case $z=50$ (cf. Eq.~(\ref{eq:schulz})).}
\label{fig:schulz50}
\end{figure}

In this work we consider the case in which the species labels $\sigma$
are drawn stochastically from a (parental) distribution of the Schulz form
\cite{Schulz1939}:

\be
f\p(\sigma)=\frac{1}{z!}\left(\frac{z+1}{\bar{\sigma}}\right)^{z+1}\sigma^z\exp\left[-\left(\frac{z+1}{\bar{\sigma}}\right)\sigma\right]\:,
\label{eq:schulz}
\ee
with a mean species label $\bar{\sigma}\equiv 1$. We have elected to study the case $z=50$,
corresponding to a moderate degree of polydispersity: the standard
deviation of the species label is $\delta\equiv 1/\sqrt{z+1}\approx 14\%$
of the mean. The form of the distribution is shown in
Fig.~\ref{fig:schulz50}. Although our motivation for employing the
Schulz distribution is primarily ad-hoc, we note that it has been found
to fairly accurately describe the polydispersity of some polymeric
systems \cite{McDonnell2003}.

In the simulations reported below, the distribution $f\p(\sigma)$ was
limited for computational convenience to the range $0.5<\sigma<1.4$,
and renormalized appropriately. Imposition of lower and upper cutoffs,
in the tails of the parent distribution, obviates the need to
determine the associated portions of the chemical potential
distribution $\mu(\sigma)$ conjugate to $\rho\p(\sigma)$ -- a task that
can be numerically fraught when the parent density is very small. It
should be noted, however, that depending on the form of the parent, the
upper cutoff can have drastic effects on the
phase behaviour, even when it is located far into the tail of the
distribution \cite{wilding2005b}.

\subsection{Methodology}

We have studied the phase behaviour of the PLG model,
Eq.~\ref{eq:hamiltonian}, under conditions of fixed polydispersity via
Monte Carlo simulations within the grand canonical ensemble (GCE). Our
Metropolis algorithm invokes three types of lattice operations:
particle deletions, particle insertions, and variation of a particle's
species label $\sigma$. The latter quantity is treated as a continuous
variable in the permitted range $0.5<\sigma<1.4$; however,
distributions defined on $\sigma$, such as the instantaneous density
$\rho(\sigma)$, and the chemical potential $\mu(\sigma)$, are
represented as histograms formed by discretising this range into $100$
bins.  The simulation results presented below pertain to periodic cubic
systems of linear dimension $L=10,20,30,40$ lattice units. Further
details concerning general aspects of the simulation of polydisperse
fluids within the GCE, as well as the structure, storage and
acquisition of data, have been presented elsewhere~\cite{wilding2002d}.

The principal observable of interest is the fluctuating form of the
instantaneous density distribution $\rho(\sigma)$.  From this we derive
the distribution $p(n)$ of the overall number density $n=\int d\sigma
\rho(\sigma)$. If one further identifies $\sigma$ with particle
diameter -- subject to the proviso that our model does not account for
any diameter dependence of the hard core repulsion -- we can consider,
as an alternative to the density $n$, a notional ``volume fraction''
$\eta=(\pi/6)\int d\sigma \sigma^3\rho(\sigma)$. 

The existence of phase coexistence at given chemical potentials is
signalled by the presence of two distinct peaks in the probability
distribution $p(n)$. In order to obtain estimates of dilution line
coexistence properties at some prescribed temperature, we employ an
accurate approach recently proposed by ourselves~\cite{buzzacchi2006}.
For a given choice of $n\p$, the method entails tuning the chemical
potential distribution $\mu(\sigma)$ together with a parameter $\xi$,
such as to simultaneously satisfy both a generalized lever rule {\em
and} an equal peak weight criterion \cite{borgs1992} for $p(n)$:

\setcounter{abc}{1}
\bea 
\label{eq:methoda}
n\p\f\p(\sigma) &=& (1-\xi)\ronesig + \xi\rtwosig \\
\addtocounter{abc}{1}
\addtocounter{equation}{-1}
r&=&1
\label{eq:methodb}
\eea 
\setcounter{abc}{0}
Here the daughter density distributions $\ronesig$ and $\rtwosig$ are assigned
by averaging only over configurations belonging to either peak of
$p(n)$. The quantity $r$ is the peak weight ratio: 
\be
r=\frac{\int_{n>n^{*}}p(n)dn}{\int_{n<n^{*}}p(n)dn}\:,
\ee
with $n^{*}$ a convenient threshold density intermediate between vapor
and liquid densities, which we take to be the location of the minimum
in $p(n)$. The tuning of $\mu(\sigma)$ and $\xi$ necessary to
simultaneously satisfy Eqs.~(\ref{eq:methoda}) and (\ref{eq:methodb})
can be efficiently achieved by histogram extrapolation techniques
\cite{Ferrenberg1989}, given an initial estimate for $\mu(\sigma)$
which is provided by an iterative refinement technique, as described
elsewhere \cite{wilding2003a}. 

The value of $\xi$ resulting from the application of the above
procedure is the desired fractional volume of the liquid phase at the
nominated value of $n\p$. Cloud points are determined as the value of
$n\p$ at which $\xi$ first reaches zero (vapor branch) or unity (liquid
branch), while shadow points are given by the density of the coexisting
incipient daughter phase, which may be simply read off from the
appropriate peak position in the cloud point form of $p(n)$ and
$p(\eta)$. It should be pointed out that the finite-size corrections to
estimates of coexistence properties obtained using the equal peak
weight criterion for $p(n)$ are exponentially small in the system size
\cite{buzzacchi2006}. 

In order to obtain the phase behaviour of our model system, we scanned
the dilution line for a selection of fixed temperatures. We started by
setting $T=T_c$, the critical temperature (known from a brief
preliminary study \cite{buzzacchi2006}), and tracked the locus of the
dilution line in a stepwise fashion. This tracking procedure must be
bootstrapped with knowledge of the form of $\mu(\sigma)$ at some
initial point on the dilution line.  A suitable estimate was obtained,
for a point near the critical density, by means of the iterative
refinement procedure \cite{wilding2003a}, in conjunction with the equal
peak weight criterion for $p(n)$ discussed above.  Simulation data
accumulated at this near-critical state point was then extrapolated to
a lower, but nearby density $n\p$ by means of histogram reweighting,
thus providing an estimate of the corresponding form of $\mu(\sigma)$.
The latter was employed in a new simulation, the results of which were
similarly extrapolated to a still lower value of $n\p$. Iterating this
procedure thus enabled the systematic tracing of the whole dilution
line and the identification of cloud points. Histogram extrapolation
further permitted a determination of dilution line properties at
adjacent temperatures, thereby facilitating a systematic determination
of the phase behaviour in the $n\p-T$ plane, including special loci
such as the cloud curve and the critical isochore ($n\p=n\p_c$).
Implementation of multicanonical preweighting techniques
\cite{berg1992} at each coexistence state point ensured adequate
sampling of the coexisting phases in cases where they are separated by
a large interfacial free energy barrier. A fuller account of this
latter procedure has been presented previously~\cite{wilding2001}.

\subsection{Results}

The techniques outlined above were used to obtain accurate estimates of
key features of the phase diagram of the PLG, namely the cloud
and shadow curves, and the coexistence properties on the critical
isochore.

\begin{figure}[h]
\includegraphics[type=eps,ext=.eps,read=.eps,width=0.95\columnwidth,clip=true]{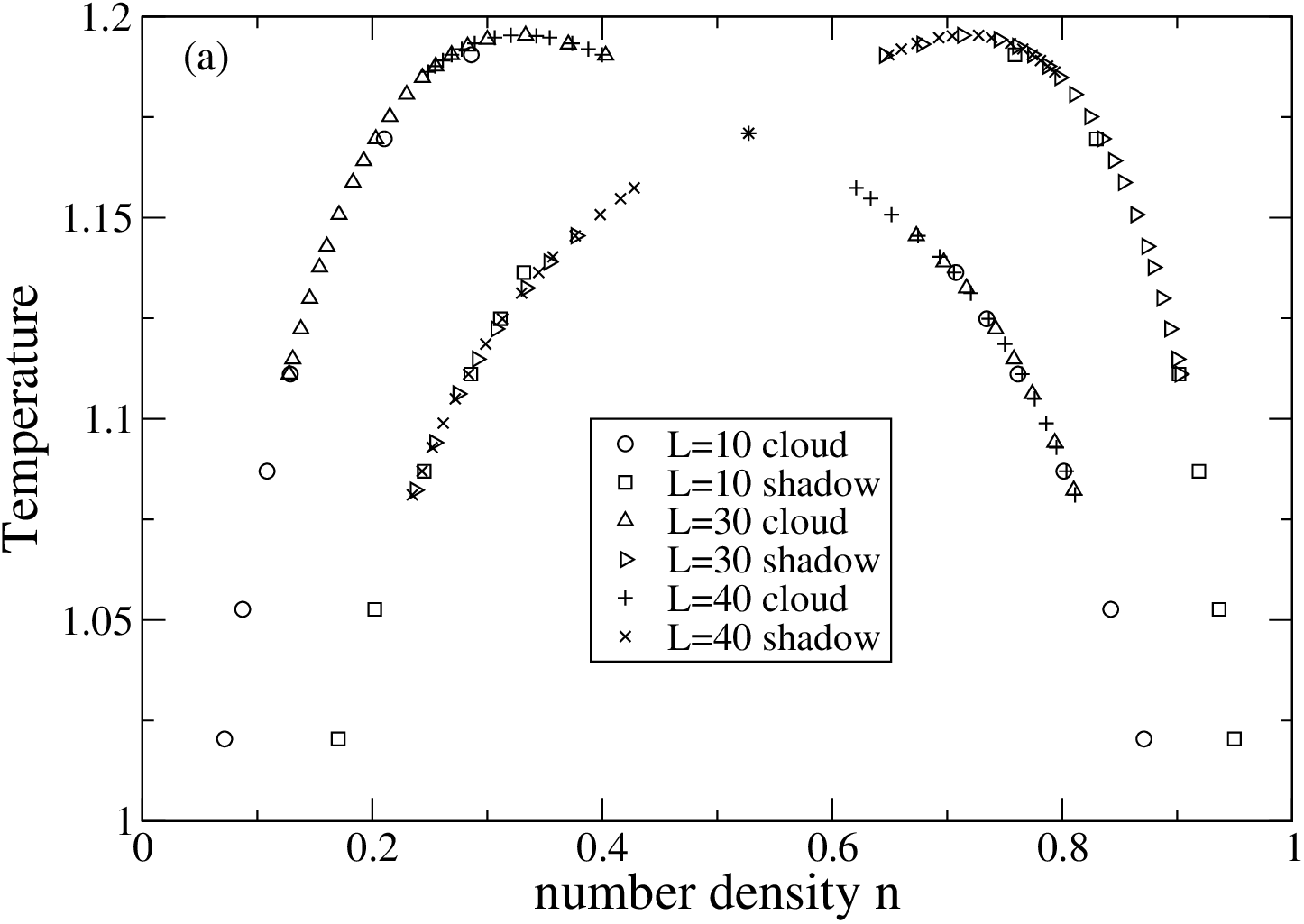}
\includegraphics[type=eps,ext=.eps,read=.eps,width=0.95\columnwidth,clip=true]{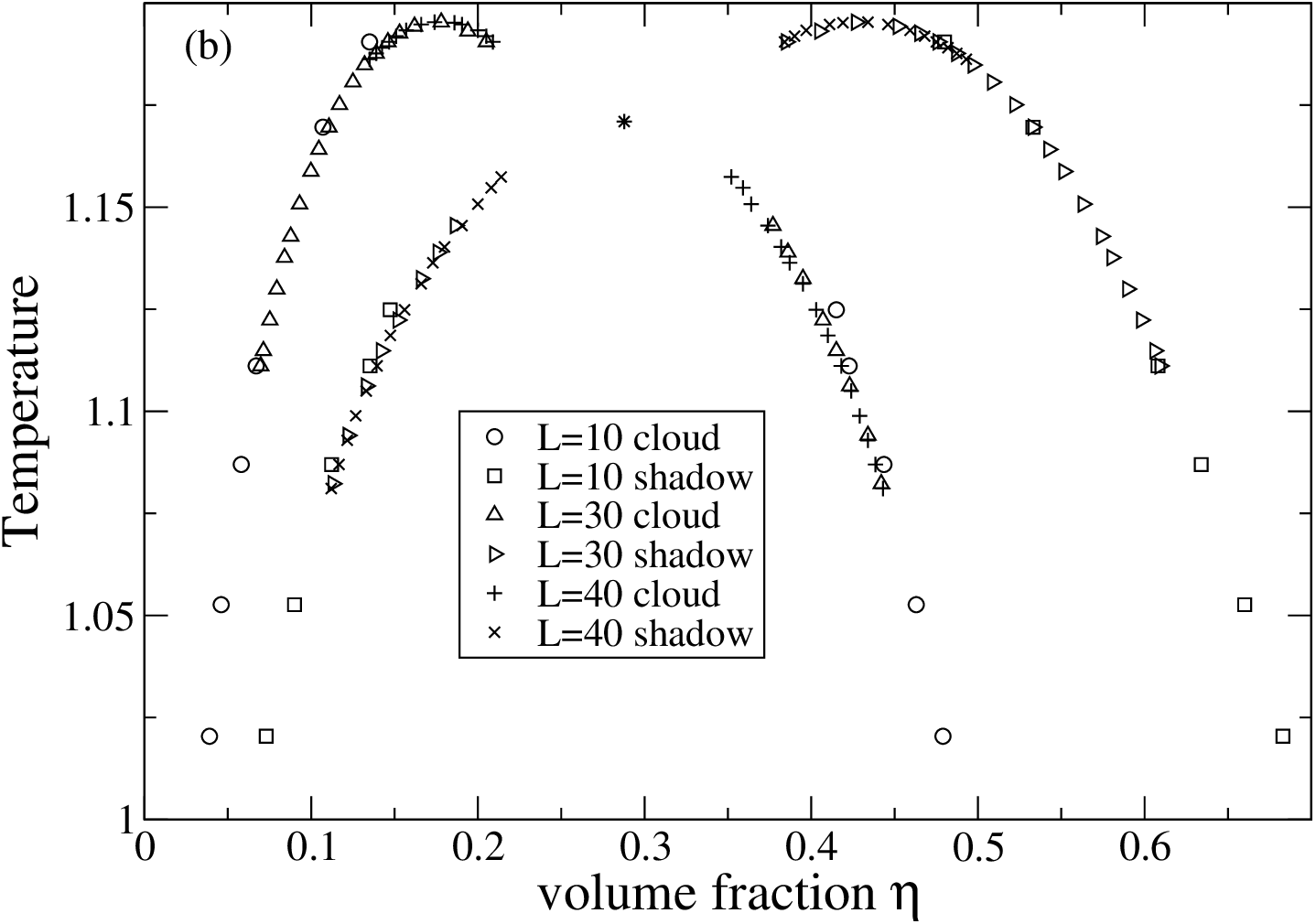}

\caption{Estimates of the cloud and shadow curves of the PLG model for
the Schulz parent distribution of Eq.~\ref{eq:schulz}, obtained as
described in the text. Data are shown for three system sizes. {\bf (a)}
Density-temperature plane. {\bf (b)} Volume fraction-temperature plane.
In each case the estimated critical point is marked by a star
($\star$).}

\label{fig:cl_sh}
\end{figure}

The liquid-vapor critical point has been located approximately in  a
previous brief study of the PLG \cite{buzzacchi2006}, which found
$(n\p_c,T_c)=(0.521,1.171)$ in reduced units. This is to be compared
with the critical parameters of the monodisperse (Ising) lattice-gas
$(0.5,1.127955)$ \cite{Luijten1995}. Thus the inclusion of
polydispersity of the form (\ref{eq:hamiltonian}) is seen to raise both
the critical temperature and the critical density. Moreover we find
that it splits the liquid-vapor binodal into well-separated cloud and
shadow curves (cf.\ Fig.~\ref{fig:cl_sh}) in a manner similar to that
occurring in continuum fluid models for which polydispersity affects
the interparticle interaction strength as well as its range
\cite{wilding2005b,Wilding2006}. (When {\em only} the range is
polydisperse, a scale symmetry forces the critical point to be near the
top of cloud and
shadow~\cite{sollich07,bellier-Castella2000,wilding2004a,wilding2004b}.)
As a consequence, the critical point lies below the maximum of the
cloud curve and phase coexistence can be observed even at $T=T_c$,
provided that $n\p<n\p_c$. Note that our procedure for cloud/shadow
curve determination breaks down in the vicinity of the critical point,
because the two peaks in $p(n)$ overlap in a finite-sized
system. This is the reason for absence of data near criticality in
Fig.~\ref{fig:cl_sh}.

\begin{figure}[h]
\includegraphics[type=eps,ext=.eps,read=.eps,width=0.9\columnwidth,clip=true]{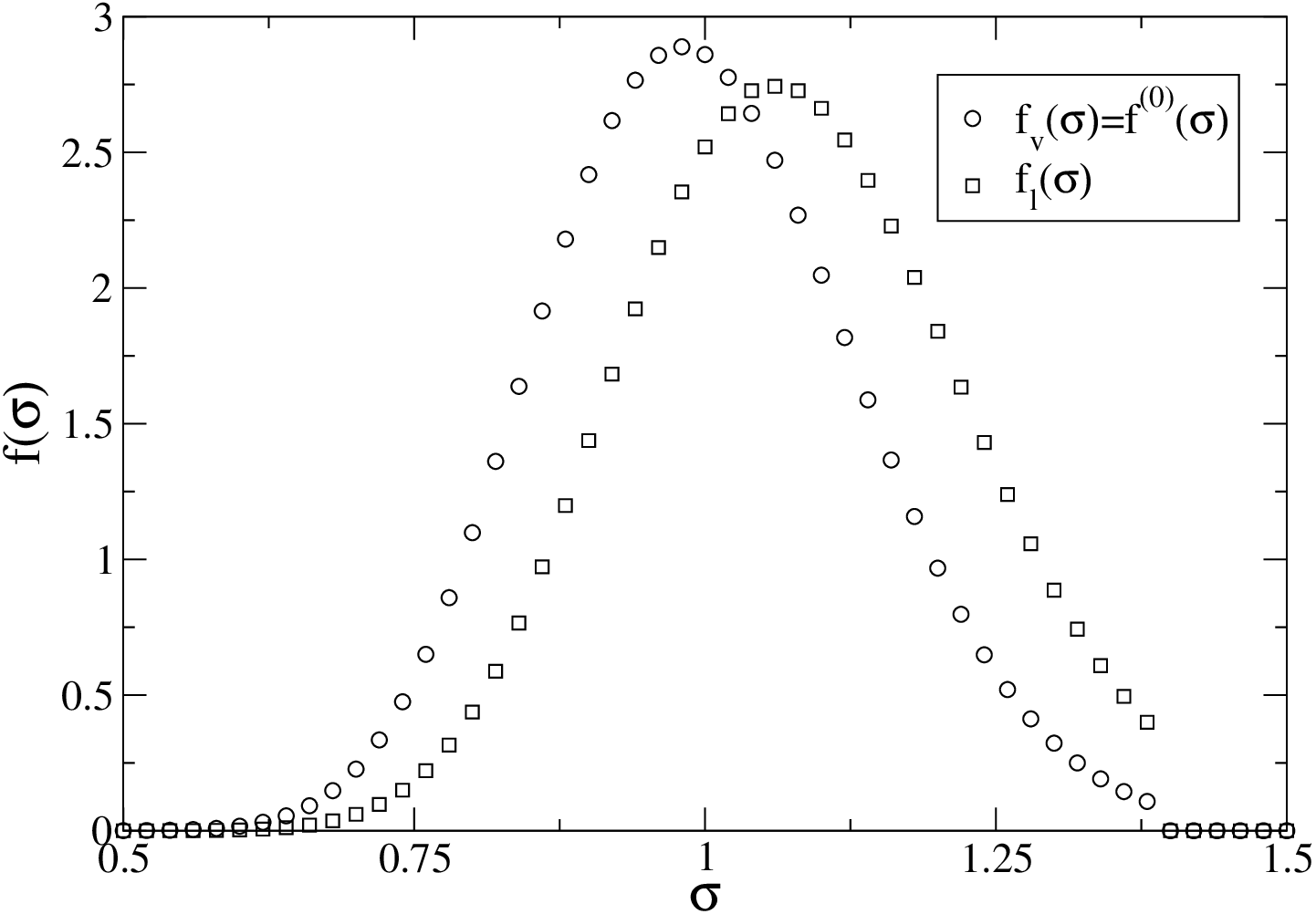}
\caption{Normalized daughter distributions $f_v(\sigma)$ and $f_l(\sigma)$ for the vapor ($v$) and liquid
($l$) phases as obtained for the $L=30$ system on the cloud curve at $n\p=0.1275, T=1.111$, as
described in the text.}
\label{fig:daughters}
\end{figure}

Fig.~\ref{fig:daughters} shows the normalized forms of the vapor and
liquid daughter phase distributions for a point on the cloud curve well
away from criticality. One sees that in keeping with previous findings
for off-lattice models \cite{wilding2005b,Wilding2006}, there is
considerable fractionation. Specifically, the number of particles in
the liquid phase having a large value of $\sigma$ is strongly enhanced
relative to the vapor phase. This enhancement arises because particles
having larger $\sigma$ interact more strongly than those of small
$\sigma$, thereby yielding a substantial free energy gain of the liquid
(with its larger number of nearest neighbor particle contacts) over the
vapor. One further sees that as a result of this enhancement, the form of
the liquid daughter phase distribution is strongly affected by the
presence of the cutoff. The issues surrounding such ``cutoff effects''
and their implications for phase behaviour have recently been examined
in detail in refs.~\cite{wilding2005b,Wilding2006}.

\begin{figure}[h]
\includegraphics[type=eps,ext=.eps,read=.eps,width=0.95\columnwidth,clip=true]{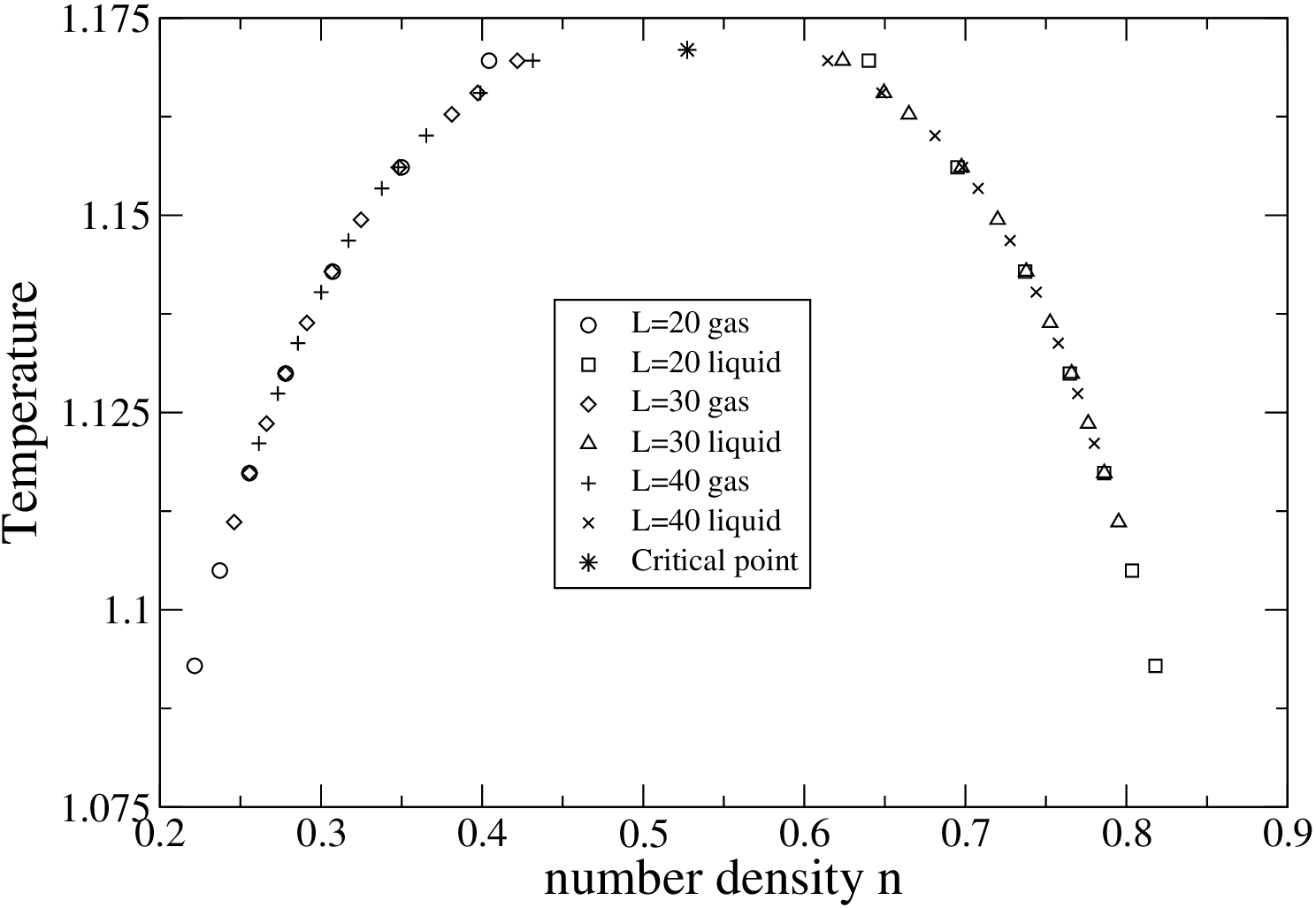}

\caption{Estimates of the coexistence curve for which the parent
density $n\p$ is fixed to its critical point value ie. $n\p=n\p_c$. Data
is shown for three system sizes. The marked densities at each
temperature derive from the peak positions in the corresponding form of
$p(n)$. Clear finite-size effects are visible in the vicinity of the
critical point.}

\label{fig:critisochor}
\end{figure}

Finally, in this section, we have determined the coexistence curve
corresponding to a parent density which is fixed at its critical value
$n\p_c=0.521$. This curve is presented in Fig.~\ref{fig:critisochor}.
One notes that its general shape is similar to the standard binodal that one
finds in a monodisperse fluid. This is because the fractional volumes
of the two phases are approximately (though not strictly) equal on this
coexistence curve. By contrast, on the cloud curve the fractional
volume of one phase is infinitesimal, leading to much stronger
fractionation effects.

\section{Conclusions}
\label{sec:concs}

In this paper, we have introduced a lattice-gas model for a
polydisperse fluid and shown that it exhibits qualitatively similar
liquid-vapor phase behaviour to realistic continuum fluid models in
which polydispersity affects both the size and the strength of the
particle interactions \cite{wilding2005b,Wilding2006}. In tests we find that the
coexistence properties of the PLG can be determined with a
computational effort that is smaller by a factor of $1-2$ orders of
magnitude compared to that for an off-lattice system having a similar
number of particles at equivalent state points such as criticality. 
This bodes well for the prospects of harnessing the PLG model to
elucidate how coexistence properties are affected by factors such as
the form of the parent distribution, the choice of the large-$\sigma$
cutoff to the parent distribution, and the $\sigma$-dependence of the
interaction strength (which is controlled in the PLG by the exponent
$\alpha$ in Eq.~\ref{eq:hamiltonian}). Additionally the model, and the
results we have provided, may prove useful as a computationally
efficient test bed for anyone wishing to ``tool-up'' for simulation
studies of polydispersity on other systems.

Finally we note that experiments have been performed on polydisperse
polymers which report that the critical exponents for the critical
coexistence curve are Fisher-renormalized with respect to the standard
Ising values \cite{kita1997a}. The present model may provide a useful
platform for elucidating these intriguing findings; this is something which we
hope to do in future work.

\end{document}